\pdfoutput=1
\documentclass{article}
\usepackage[utf8]{inputenc}
\usepackage{url}
\usepackage{authblk}
\usepackage[T1]{fontenc}
\usepackage{lmodern}
\usepackage{amsfonts}
\usepackage{appendix}

\usepackage{tikz,pgfplots}
\pgfplotsset{compat=1.12}
\usepgfplotslibrary{dateplot}

\title{Green NFTs: A Study on the Environmental Impact of Cryptoart Technologies}

\date{May 2021}

\usepackage{natbib}
\usepackage{graphicx}
\usepackage{fixltx2e}
\usepackage{mathtools}
\usepackage[normalem]{ulem}
\useunder{\uline}{\ul}{}
\usepackage{outlines}

\begin{document}




\author{Samuele Marro\thanks{marrosamuele@gmail.com (equal contribution)} }
\author{Luca Donno\thanks{donnoluca@outlook.com (equal contribution)}}
\affil{University of Bologna}
\maketitle


\begin{abstract}
    We introduce a model of greenhouse gas emissions due to on-chain activity on Ethereum, focusing on cryptoart. We also estimate the impact of individual transactions on the environment, both before and after the London hard fork. We find that with the current fee mechanism, spending one dollar on transaction fees corresponds to emitting at least the equivalent of 1.305 kilograms of CO\textsubscript{2}. We also describe several techniques to reduce cryptoart emissions, both in the short and long term.
\end{abstract}

\section{Introduction}

In the last year, there has been an exponential growth of cryptoart, a new blockchain-based art form. After the publication of several articles that aimed to raise awareness about its environmental impact \citep{akten2020unreasonable} \citep{lemercier2021joanie}, there has also been a growing interest in developing solutions to reduce greenhouse gas (GHG) emissions due to cryptoart. A range of proposals have been advanced, from reducing gas usage \citep{pipkin2021here} to switching to Proof of Stake blockchains \citep{wintermeyer2021climate} to carbon offsets \citep{kahn2021how}. Others have even asserted that cryptoart has no impact on carbon emissions \citep{superrare2021no} \citep{mattei2021should}.
We argue that a model of the environmental impact of cryptoart is fundamental to properly evaluate the effectiveness of potential solutions.
We therefore introduce a model to compute the GHG emissions of cryptoart hosted on the Ethereum blockchain. We also analyze the impact of a single transaction, focusing on both the current fee mechanism and the soon to be implemented EIP-1559. By estimating the geographical distribution of Ether miners and their cost breakdown, we find that spending one dollar in transaction fees leads to additional GHG emissions amounting to 1.305 kgCO\textsubscript{2}eq.
Finally, we illustrate different solutions that would reduce the environmental impact of cryptoart, focusing on both individual and collective actions.
A summary of our key conclusions can be found in Section  \protect\ref{sec:keyTakeaways}.

\section{Background}

\subsection{Cryptoart}
Cryptoart is a category of art based on blockchain technology. Artworks are stored as Non-Fungible Tokens (NFTs), unique digital assets whose ownership and scarcity can be verified in a decentralized and trustless manner.
The actions of creating (also known as "minting"), bidding, buying and transferring NFTs are performed by executing transactions on-chain.
Some of the most known cryptoart platforms, such as OpenSea\footnote{Website: opensea.io}, Nifty Gateway\footnote{Website: niftygateway.com} and SuperRare\footnote{Website: superrare.co} are currently hosted on the Ethereum network.

\subsection{Block production}

In order for a transaction to be accepted by the network, it must be included in a block. A block is a data structure containing a collection of transactions.
Protocols define an entity, called validator, which appends new blocks to the blockchain.
Since blocks are limited in size and frequency, validators prioritize inserting transactions that maximize their revenue. Users therefore pay fees to be included in new blocks. A transaction with a higher fee will be included sooner.




\subsection{Proof of Work}

Proof of Work (PoW) blockchains use a consensus algorithm where blocks are validated by solving an energy intensive computational puzzle.
The protocol defines a value, block difficulty, which is adjusted in order to maintain a fixed block frequency. The energy consumption required to validate (or mine) a block is proportional to the block difficulty.

The computational power of a miner is measured by its hash rate. The difficulty of a block is proportional to the total hash rate of the network.
The probability that a miner will mine a block is proportional to the percentage of their own hash rate compared to that of the network.
Given a certain difficulty, both full and empty blocks require approximately the same amount of energy to be mined.

While Bitcoin is mined on application-specific integrated circuits (ASICs), Ethereum is relatively ASIC-resistant and is therefore mined on graphical processing units (GPUs).

\subsection{EIP-1559}
EIP-1559 is an Ethereum update that will take place in July with the London hard fork.

The current fee mechanism is based on first-price auctions.
A transaction requires a certain amount of block space, called gas, which is proportional to its complexity. Each user chooses the price paid per unit of gas. Miners prioritize transactions with the highest gas price; therefore, users have to estimate the minimum gas price required to be included.
Since users do not have perfect information on other bids, this price tends to be overestimated.

EIP-1559 solves this problem by introducing a protocol-defined gas price, \texttt{BASEFEE}, which must be paid for a transaction to be included in the block.
EIP-1559 defines a target block size $T$ and a maximum block size $2T$. If the previous block was larger than $T$, \texttt{BASEFEE} is increased, and vice versa.

\texttt{BASEFEE} is computed as follows:
$$B_{n+1}=B_{n} \left (1+\frac{1}{8}\frac{B_n-T}{T} \right )$$
where $B_n$ is the \texttt{BASEFEE} of the $n$-th block.

The key difference between EIP-1559 and the current fee mechanism is that the transaction fee is burned. To incentivize miners to include transactions, users pay an additional \texttt{MINER\_TIP}, which only covers execution costs.

If, at any point in time, the total size of pending transactions is higher than $2T$, this new fee mechanism degrades to a first-price auction using \texttt{MINER\_TIP}.

\citet{roughgarden2020transaction} predicts that EIP-1559 will not affect the gas price.

\section{Miners}

In this section, we show that, under reasonable assumptions, the global mining cost of a PoW blockchain is approximately equal to the global mining revenue.

We treat miners as rational agents that aim to increase their profits. 
A rational agent would not undertake an activity which would result in a ongoing loss in doing so.
This approach ignores other incentives that might motivate a miner, such as the desire for a more decentralized blockchain.

Miners have two main sources of revenue:
\begin{outline}
    \1 The block reward, which consists of:
        \2 The fixed block reward;
        \2 The uncle block reward\footnote{Uncle block rewards and uncle block inclusion rewards \citep{wood2014ethereum} are mechanisms to promote decentralization. They are designed to reward miners that successfully mined a block which was not validated by the network. \label{uncleBlock}};
        \2 The uncle block inclusion reward;
    \1 Transaction fees.
\end{outline}

and three main costs:

\begin{itemize}
    \item Electricity cost;
    \item Hardware cost;
    \item Other types of costs, which we assume to be negligible.
\end{itemize}

The probability of a miner collecting the block reward and the transaction fees is equal to their hash rate divided by the global hash rate. More formally, we model the expected revenue of a miner $k$ from a block as:

\begin{equation}
    \mathbb{E} \left [R_k \right ] = \left (R_{block} + R_{fees} \right ) \frac{H_k}{ \sum_i H_i}
\end{equation}
where $R_{block}$ is the revenue due to the block reward, $R_{fees}$ is the revenue from transaction fees and $H_i$ is the hash rate of the $i$-th miner.

Since miners have a negligible impact on $R_{block}$ and $R_{fees}$, the only way to increase their revenue is to increase their relative hash rate. We treat variations in the per-unit cost of hardware and electricity due to economy of scale effects as negligible.
If GPUs are being already used at maximum performance, a higher hash rate can only be achieved by acquiring more GPUs. Since the hardware cost of a typical mining setup is dominated by GPUs, doubling the number of GPUs also roughly doubles the hardware cost. Similarly, since the power requirements are dominated by GPUs, doubling their number would also approximately double the electricity cost.
GPUs, however, have a limited lifetime before becoming outdated or unusable. We therefore treat GPUs not as a one-time cost but as a recurring one.

More formally, we model the hash rate of a miner as directly proportional to both hardware and electricity costs.

Since Ether can be mined using consumer-grade hardware and with minimal technical skills, new miners can easily enter the market.
We therefore model Ether mining as a competitive virtual commodity market, a market in which agents reach an equilibrium where the marginal revenue is equal to the marginal cost. 
In other words, the revenue of mining for an additional day is equal to the cost.
The total mining revenue during a certain period of time is therefore equal to the total mining cost:
\begin{equation}
    R_{tot} = C_{tot}
\end{equation}

Our findings match the results obtained by \citet{hayes2017cryptocurrency}, who applied a similar model to estimate the GHG impact of Bitcoin.

Our model assumes that miners can quickly adapt their costs to market fluctuations. This is possible due to the existence of alternative sources of revenue (e.g. mining on other blockchains). If the profitability of Ether mining decreases, some miners will redirect their computational power towards other tasks, and vice versa.
For the purpose of this paper, we ignore variations in the profitability of such activities.

\section{GHG Impact of Proof of Work}

In this section, we show that the GHG impact of a PoW blockchain is directly proportional to its global mining revenue.

First, we assume that variations in mining revenue do not significantly affect the distribution of energy sources, as well as of the hardware used.

Therefore, we treat the GHG impact of hardware and electricity as directly proportional to their costs.
We model the GHG emissions due to mining as follows:
\begin{equation}
    E_{tot} = \alpha_{hw} C_{hw} + \alpha_{el} C_{el}
\end{equation}
where $E_{tot}$ are the global mining emissions (in kgCO\textsubscript{2}eq) of mining, $C_{hw}$ is the global mining cost due to hardware, $C_{el}$ is the global mining cost due to electricity, and $\alpha_{hw}$ and $\alpha_{el}$ are constants that represent the kgCO\textsubscript{2}eq emitted per dollar spent respectively on hardware and electricity.

Assuming that variations in mining revenue do not meaningfully influence the ratio between hardware and electricity costs, we rewrite $C_{hw}$ and $C_{el}$ as follows:
\begin{equation}
    C_{hw} = \beta C_{tot}
\end{equation}
\begin{equation}
    C_{el} = (1 - \beta) C_{tot}
\end{equation}
where $C_{tot} = C_{hw} + C_{el}$ and $0 \leq \beta \leq 1$ is a constant.

$E_{tot}$ can be therefore rewritten as:

\begin{equation}
    E_{tot} = \alpha_{hw} \beta C_{tot} + \alpha_{el} (1 - \beta) C_{tot} = \alpha_{tot} C_{tot} = \alpha_{tot} R_{tot}
\end{equation}
where $\alpha_{tot} = \alpha_{hw} \beta + \alpha_{el} (1 - \beta)$ is a constant measuring the kgCO\textsubscript{2}eq emitted per dollar spent on mining.

\section{Estimating $\alpha_{tot}$}
In this section, we estimate the kgCO\textsubscript{2}eq emitted per dollar earned by Ether miners.

We compute a lower bound of $\alpha_{tot}$ by considering only two sources of GHG emissions, namely GPU production and electricity usage.
In order to do so:
\begin{itemize}
    \item We choose a representative mining GPU;
    \item We estimate the geographical distribution of Ether miners;
    \item We estimate the weighted average kgCO\textsubscript{2}eq emitted per dollar spent on electricity;
    \item We estimate the average lifetime of a GPU and the ratio of $C_{el}$ and $C_{hw}$ compared to $C_{tot}$ in that time frame;
    \item We use our gathered data to compute an estimate of $\alpha_{tot}$.
\end{itemize}

\subsection{Representative GPU}

For the sake of modelling the role of GPUs in mining emissions, we use the AMD RX 590 as a representative GPU. As of May 2021, the RX 590 is the GPU with the highest Ethereum hash rate per Watt-hour \cite{minerstat2021gpu}. The relevant specifications of an RX 590 are outlined in table \ref{tab:standardGpu}.

Following \citet{vries2018bitcoin}, we estimate a maximum GPU lifespan of 2 years.
\citet{ardente2015environmental} estimate that producing a GPU has a GHG impact of 54 kgCO\textsubscript{2}eq.

\begin{table}[]
\centering
\begin{tabular}{|l|c|}
\hline
Name              & AMD RX 590 \\ \hline
Launch quarter   & Q4 2018     \\ \hline
Current average price      &    USD 650      \\ \hline
Hash rate          &    27.31 MH/s      \\ \hline
Power consumption &    163 W     \\ \hline
\end{tabular}
\caption{Specifications of an AMD RX 590 GPU. Since November 2020, a spike in demand has led to several shortages and an increase in average prices from USD 450 to USD 850. We therefore use an estimate of the average price in the last 6 months. Sources: \citet{amd2018rx}, \citet{minerstat2021gpu} \citet{camel2021rx}.}
\label{tab:standardGpu}
\end{table}

\subsection{Geographical Distribution of Miners}
While the approximate global hash rate of the Ethereum network is known, the inherent anonymity and decentralization of its protocol prevent accurate estimates of the geographical distribution of its nodes.

A naive approach would be to count the number of Ethereum nodes in each country. This technique, however, ignores potential differences in node hash rate between different countries.

A second approach involves using data on mining pools. Mining pools represent a large portion of the global hash rate \citep{lin2021measuring} and often provide information on their hash rate share. However, several large mining pools are based in multiple countries and do not provide information regarding their internal geographical distribution. Moreover, if the increased payout and reliability outweigh the revenue loss due to connection delays, a miner might join a pool in a different country or continent.

We therefore follow a different strategy: we use the data collected by \citet{silva2020impact} on block observation as a proxy of the true geographical distribution of Ether miners. In April 2019, the authors collected statistics on which of four nodes (located in Western Europe, Central Europe, North America and East Asia) was the first to receive updates on a newly mined block. While during the last two years there might have been shifts in the geographical distribution, we argue that, due to the relative stability of Ether mining economics\footnote{Compare with Bitcoin, which saw a shift towards industrial, ASIC-based large scale mining \citep{taylor2017evolution}.}, such estimates are still adequate for our purposes.
In the context of impact estimation, we treat Western and Central Europe as the same region. We also consider the role of other regions negligible.
We compute a weighted average of kgCO\textsubscript{2}eq per dollar based on average industrial electricity prices and carbon-intensity-per-kilowatthour (CIPK), as outlined in table \ref{tab:electricity}.

\begin{table}[]
\centering
\begin{tabular}{|l|l|l|l|l|}
\hline
\textbf{Region}        & \textbf{Hash rate \%} & \textbf{\$/kWh} & \textbf{kgCO2eq/kWh} & \textbf{kgCO2eq/\$} \\ \hline
Europe        & 50          &   0.1419     &      0.230     &   1.621  \\ \hline
East Asia          & 38          &    0.0916    &      0.582     &   6.354  \\ \hline
North America & 12          &    0.0815    &      0.331     &  4.061      \\ \hline
Overall &  \textasciitilde 100      &   0.1155     &   0.376     &    3.712 \\ \hline 
\end{tabular}

\caption{Industrial electricity prices and CIPK by region. We use average electricity prices in the European Union \citep{rademaekers2020study}, People's Republic of China \citep{escap2018sustainable} and the United States \citep{eia2021monthly}. Source for CIPK data: \citet{iea2020composition}.}
\label{tab:electricity}
\end{table}

\subsection{Global estimate}

We conclude that, over a 2-year period, a single GPU will cost 980 dollars and have an impact of 1279 kgCO\textsubscript{2}eq (see table \ref{tab:alpham}).
Our estimate of $\alpha_{tot}$ is therefore 1.305 kgCO\textsubscript{2}eq/\$. 
In order to put this value into context, we apply it to the current Ethereum data. From April 29, 2021, to May 5, the average global daily revenue for Ethereum miners was 58.91 millions of dollars \citep{etherchain2021mining} \citep{etherchain2021value}, equivalent to a daily emission of 76.89 ktCO\textsubscript{2}eq or an annual emission of 28.06 MtCO\textsubscript{2}eq. The latter value is compatible with the 19.64 MtCO\textsubscript{2}eq figure estimated by the Ethereum Energy Consumption Index over the same period \citep{vries2021ethereum}.


\begin{table}[]
\centering
\begin{tabular}{|l|l|l|}
\hline
\textbf{Name}        & \textbf{Cost (USD)} & \textbf{GHG emissions (kgCO\textsubscript{2}eq)} \\ \hline
Hardware    & 650          & 54         \\ \hline
Electricity & 330         &  1225        \\ \hline 
Total       & 980         & 1279         \\ \hline
\end{tabular}
\caption{Final breakdown of costs and emissions for one RX 590 GPU over 2 years.}
\label{tab:alpham}
\end{table}

\section{Individual Impact}
\label{sec:individualImpact}

While estimating the total emissions of a network can be useful to compare entire blockchains, we also model the impact of a single transaction, both with the current fee mechanism and with EIP-1559.
In order to isolate the effects of individual transactions, we assume that both the aggregate real value of transactions on Ethereum and the velocity of Ether are constant for a given period of time.

\subsection{First-Price Auction}

Suppose that a user successfully executes a transaction that requires $G$ gas at a price (in Gwei, equivalent to $10^{-9}$ Ether) of $P_{gas}$ per unit of gas.
We identify three main effects of this action:
\begin{enumerate}
    \item The user paid their fee to a miner, thereby directly increasing the global mining revenue; \label{feeToMiner}
    \item The user caused other transactions amounting to $G$ gas to be left out of the block, which are then\footnote{Due to the specifications of the Ethereum protocol, it is not possible to reduce the gas price of a submitted transaction.}: \label{leftOut}
    \begin{enumerate}
        \item Accepted later at the same price; \label{acceptedLater}
        \item Accepted after their gas price is increased; \label{reEnterWithHigher}
        \item Never accepted;\label{neverAccepted}
    \end{enumerate}
    \item The user increased estimates of the gas price, leading other users to bid more in the following blocks. \label{gasPriceIncrease} \label{ethereumPriceIncrease}
\end{enumerate}

On May 1, out of more than 1.1 billion total Ethereum transactions \citep{blockchair2021ethereum}, roughly 190 thousand (<0.02\%) were still pending, the majority of which were accepted within 24 hours \citep{etherscan2021ethereum}. We therefore treat case \ref{neverAccepted} as negligible.

We compute a lower bound of the increase in miner revenue by considering only cases \ref{feeToMiner}, \ref{acceptedLater} and \ref{reEnterWithHigher}. This approach ignores the long-term effects of a transaction, therefore providing a conservative estimate of the true impact.

Specifically, we model the lower bound of the overall increase in miner revenue (in dollars) $\Delta R$ as:
\begin{equation}
    \Delta R \geq P_{Gwei} G P_{gas} + P_{Gwei} \sum_i  g_i \left (p_i' - p_i \right ) \label{grossMinerRevenueTransaction}
\end{equation}
where $G= \sum_i g_i$, $P_{Gwei}$ is the price of one Gwei, and $g_i$, $p_i$ and $p_i'$ are respectively the gas amount, the old gas price and the new gas price of the $i$-th transaction that was excluded.
For case \ref{acceptedLater}, $p_i' = p_i$, which means that the direct revenue impact of delaying such transactions is zero. We compute a looser lower bound by ignoring case \ref{reEnterWithHigher}\footnote{We analyzed 20k transactions executed on May 3, 2021, finding that case \ref{reEnterWithHigher} was responsible for an increase of less than 2\% in transaction revenue. \citep{etherscan2021transactions}}.

We therefore simplify inequality \ref{grossMinerRevenueTransaction} by rewriting it as:
\begin{equation}
    \Delta R \geq G P_{Gwei} P_{gas}
\end{equation}
In other words, the lower bound $\Delta E$ of the GHG impact of a transaction is:
\begin{equation}
    \Delta E \geq \alpha_{tot} G P_{Gwei} P_{gas}
\end{equation}

\subsection{EIP-1559}


With EIP-1559, since the \texttt{MINER\_TIP} only covers the execution costs, the total mining revenue is dominated by the block reward.
Therefore, the only non-negligible impact that a transaction has on mining revenue is the deflationary effect of burning \texttt{BASEFEE} Ether.

Let $s_t$ be the rate of growth of the Ether circulating supply after the $t$-th block has been mined:
\begin{equation}
    s_t = \frac{u_t}{S_t}
\end{equation}
where $u_t$ is the absolute variation in the number of existing Ethers and $S_t$ is the Ether circulating supply after the $t$-th block has been mined.

Since we treat the real value of transactions on Ethereum and the velocity of Ether as constant, by applying Fisher's equation for the quantity theory of money we find that the dollar value $P_{t+1}$ of one Ether after the $t+1$-th block has been mined is:
\begin{equation}
    P_{t+1} = \frac{P_t}{s_t + 1} = P_t \frac{S_t}{S_t + u_t}
\end{equation}

Let $V_t = P_t S_t$ be the total dollar value of the Ether circulating supply after the $t$-th block has been mined. Note that
\begin{equation}
    V_{t+1} = P_{t+1} S_{t+1} = P_t \frac{S_t}{S_t + u_t} (S_t + u_t) = P_t S_t = V_t
\end{equation}

In other words, $V_t$ is constant. Let $V : \forall t . \; V_t = V$. $P_t$ is therefore:
\begin{equation}
    P_t = \frac{V}{S_t}
\end{equation}

In EIP-1559, there are two factors that have opposite effects on the circulating supply. After each block is mined, $m = 2$ additional Ethers are created to reward the miners, while users burn the equivalent of $b$ dollars in transaction fees. We assume $m$ and $b$ to be constant.
$u_t$ is therefore:
\begin{equation}
    u_t = m - \frac{b}{P_t} = m - \frac{b}{V}S_t
\end{equation}

$S_{t+1}$ is then equal to:
\begin{equation}
\label{supply}
    S_{t+1} = S_t + u_t = S_t + m -\frac{b}{V}S_t
\end{equation}
while the miner revenue $R_{t+1}$ in USD after the $t+1$-th block has been mined is:
\begin{equation}
\label{reward}
    R_{t+1} = R_t +m\frac{V}{S_{t+1}}
\end{equation}

Let $S_0$ be the Ether circulating supply at a $t = 0$.

$R_t$ can be rewritten as:
\begin{align}
\begin{split}
    R_t =& R_{t-1}+\frac{mV}{S_t}= \\ =& R_{t-2}+\frac{mV}{S_{t-1}}+\frac{mV}{S_t} = \\
    &\vdotswithin{=} \notag \\
    =&R_0+\sum_{j=1}^t \frac{mV}{S_j}
\end{split}
\end{align}

Since $R_0=0$:
\begin{equation}
    R_t = mV\sum_{j=1}^t \frac{1}{S_j}
\end{equation}

$S_{t+1}$ can be rewritten as:
\begin{equation}
    S_{t+1}=S_t+m-\frac{b}{V}S_t=S_t\underbrace{(1-\frac{b}{V})}_{:=K}+m=KS_t+m
\end{equation}

which is equal to:
\begin{align}
    \begin{split}
        S_{t+1} =& KS_t+m = \\
        =& K^2S_{t-1}+mK+m = \\
        =& K^3S_{t-2}+mK^2+mK+m =\\
        &\vdotswithin{=} \notag \\
        =& K^{t+1}S_0+m\sum_{j=0}^{t}K^j
    \end{split}
\end{align}

$S_t$ is therefore equal to:
\begin{equation}
    S_t = K^tS_0+m\left(\frac{1-K^t}{1-K}\right)
\end{equation}

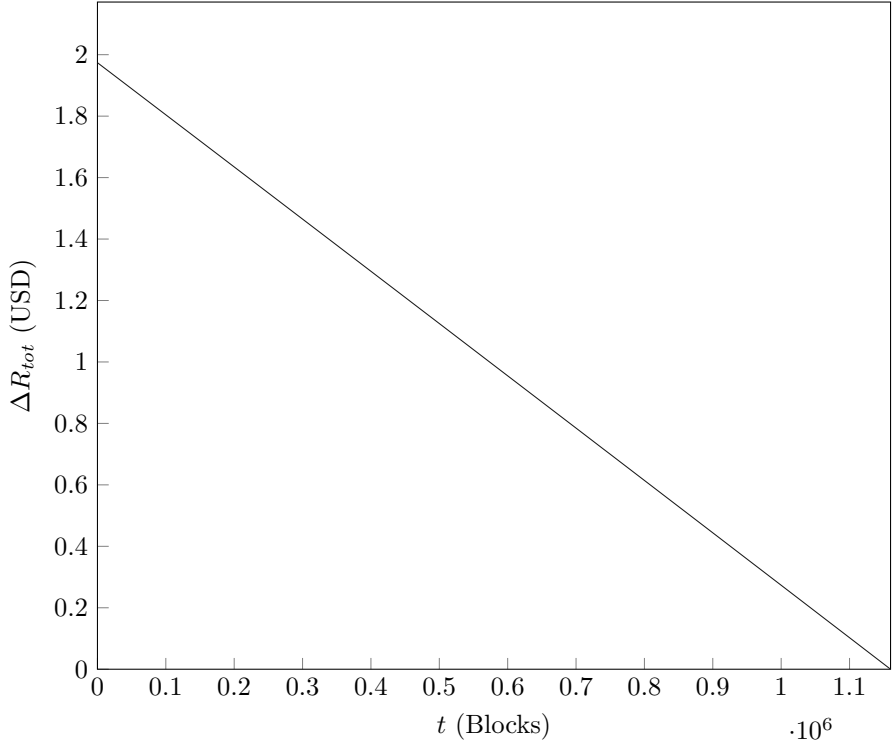
\begin{figure}[!tbhp]
    \centering
    \begin{tikzpicture}
    	\begin{axis}[width=\linewidth,
    	xmin=0,
    	xmax=1160000,
    	ymin=0,xlabel = {$t$ (Blocks)},
    	ylabel = {$\Delta R_{tot}$ (USD)},
    	tick pos=left
    	]
    	\addplot [draw=black!90!white] table [x=Block, y=Revenue, col sep = semicolon] {EIP1559Impact.csv};
    	\end{axis}
    \end{tikzpicture}
    \caption{Difference in miner revenue due to the execution of a transaction which requires a fee of 100 dollars. The x axis represents the time of execution, with $t=0$ corresponding to the first block after the introduction of EIP-1559. We set $S_0$ = 115.7M Ether and $V_0$ = 341B USD,  corresponding to the average circulating supply and total value on May 1, 2021 \citep{coingecko2021ethereum}. We use $b$ = 2650 USD, equal to the average revenue per block due to transaction fees from April 29, 2021 to May 5, 2021 \citep{etherscan2021daily}.}
    \label{fig:transactionEffect}
\end{figure}

Spending $l$ dollars on transaction fees is equivalent to removing $lP_0 = l\frac{V}{S_0}$ Ethers from the initial circulating supply.
Between the deployment of EIP-1559 (in July) and the transition to Proof of Stake (at the beginning of 2022, after which the additional environmental impact of the transaction will be negligible), only around 1.16 million blocks are expected to be mined.
Since the closed-form expression of $R_t$ (which we report in Appendix \ref{app:closedForm}) does not provide an intuition on the effect of a transaction, we instead plot the additional miner revenue $\Delta R_t$.
As shown in figure \ref{fig:transactionEffect}, the additional revenue is at least one to two orders of magnitude lower than with the current fee mechanism.

We estimate that, under our hypotheses, during the approximately 180 days between the introduction of EIP-1559 and the transition to PoS, the total miner revenue with the current fee mechanism would be 9.9 billion dollars.
Over the same period, with EIP-1559 the miner revenue would be 6.8 billion dollars, corresponding to a reduction in the GHG emissions of Ethereum by approximately a third.

\section{Solutions}

In this section, we present the most relevant solutions to reduce the GHG impact of NFTs, focusing on both individual and collective actions.

\subsection{Individual Actions}

From the model described in section \ref{sec:individualImpact}, we deduce that any action to reduce transaction fees has a beneficial impact on GHG emissions.
We therefore outline some practices to reduce the individual impact of a user.

A simple yet effective solution is to perform transactions using a lower gas price. This can be done by tolerating a longer confirmation time, executing transactions during times of the day with lower gas demand, or both.

Another way to reduce the environmental impact is by decreasing gas usage. For example, lazy minting minimizes gas requirements by treating the creation and sale of an NFT as a single transaction.

Additionally, artists often auction NFTs on-chain. Since on-chain bids require paying a transaction fee, we suggest using off-chain bids (supported for example by the Wyvern Protocol, implemented by OpenSea) to help reduce emissions.

Finally, most NFTs are sold through instant sales, where artists sell their work for a fixed price. If the artwork is particularly sought after, buyers are incentivized to use higher gas prices to outrun the others, therefore paying a much higher transaction fee than necessary. Moreover, buyers that fail to obtain a copy also pay transaction fees. In such cases, switching from instant sales to off-chain bids can significantly reduce the impact of cryptoart.

\subsection{Proof of Stake}
Proof of Stake (PoS) is a class of consensus algorithms that select and reward validators as a function of their economic stake in the network \citep{bentov2016cryptocurrencies}. Unlike PoW, the probability of creating a block in a PoS network does not depend on computational power, but rather on the staked amount that might be lost in case of fraudulent activity.
A PoS node has minimal hardware and electricity requirements, and therefore has a significantly lower environmental impact. Moreover, increasing the stake has no effect on the energy consumption of the node. Ethereum 2.0, the upcoming PoS version of Ethereum, is expected to reduce the environmental impact of the network by 99\% \citep{fairley2019ethereum}.

\subsection{Layer 2}
Layer 2 is a collective term for solutions designed to help scale applications by handling transactions off the main chain (referred to as layer 1). Generally speaking, transactions are submitted to layer 2 nodes, which batch them into groups before storing them on layer 1. 

Layer 2 technologies that are based on Ethereum have a nontrivial impact on the environment. However, layer 2 transactions require significantly less gas and therefore represent a preferable alternative to executing them directly on-chain.
The main types of layer 2 solutions are sidechains and rollups.

\subsubsection{Sidechains and Plasma}

A sidechain is a separate blockchain which runs parallel to the layer 1 and operates independently, usually with a Proof of Stake consensus algorithm. Sidechains are connected to the main chain by a two-way bridge and are often less decentralized and therefore less secure than their layer 1.

Plasma, on the other hand, is a technology that provides a way to execute transactions off-chain at a higher frequency and lower cost. It regularly stores on layer 1 information such as asset ownership, protecting it from attacks. However, in extreme cases, it is not guaranteed that the full state of Plasma can be recovered \citep{poon2017plasma}.
Sidechains and Plasma can be used together, such as in the case of Polygon, which features a hybrid protocol to mitigate the risk of information loss \citep{kanani2019matic}.

\subsubsection{Rollups}

Rollups are a class of layer 2 technologies that execute transactions outside layer 1 but store their data on-chain. This approach guarantees that, even in case of attack, it is always possible to recover all performed transactions.
The same transaction, if executed on a rollup, requires two to three orders of magnitude less gas compared to its layer 1 counterpart \citep{buterin2021incomplete}.
Rollups can be divided into zero-knowledge and optimistic rollups.

Zero-knowledge (ZK) rollups bundle transactions off-chain and generate a cryptographic proof of their validity. This proof, known as a SNARK (Succinct Non-interactive ARgument of Knowledge), is then stored on layer 1.
Being a relatively new technology, as of May 2021 the only known NFT platform hosted on ZK rollups is Immutable X\footnote{Website: immutable.com}, which is currently in closed alpha.
The main disadvantage of ZK rollups is that they do not support general computation. Because of this constraint, only basic NFT operations (e.g. minting and trading) can be executed. Extending ZK rollups is currently an active area of research.

Optimistic rollups, on the other hand, assume that all transactions are valid (hence the name "optimistic"). In optimistic rollups, anyone can submit a new transaction batch. However, if a user (also known as "fraud verifier") suspects that a batch is fraudulent, it is possible to prove it by executing the entire computation using the stored data. Fraud verifiers are usually rewarded financially at the expense of malicious users. Unlike ZK rollups, optimistic rollups support general computation, but, before withdrawing assets, users must wait a "challenge period" during which other users can claim fraudulent activity.

The two most known optimistic rollups, Optimism\footnote{Website: optimism.io} and Arbitrum\footnote{Website: arbitrum.io}, are currently in development. Optimism is expected to launch in July, while, as of May 2021, the team behind Arbitrum has not provided a launch date yet.

\subsection{Carbon Offsets}

Carbon offsets represent an alternative way to reduce the GHG impact of blockchains. By purchasing carbon offsets, a user compensates for their emissions by funding activites that have a negative GHG balance, such as planting trees or increasing the commercial viability of renewable energy.
Several initiatives that use blockchains as a tool to complement carbon markets have been introduced \citep{howson2019cryptocarbon} \citep{kahn2021how}. Moreover, some NFT platforms, such as Immutable X, compensate for their GHG emissions by automatically purchasing carbon offsets \citep{immutable2020immutable}.

Another, more indirect, method to reduce the environmental impact is to donate to climate non-profits. The cryptoart platform KnownOrigin\footnote{Website: knownorigin.io}, for example, offers an interface for artists to donate part of their revenue to sustainable causes \citep{knownorigin2021step}.

\section{Example}



To put our results in context, we use our model to estimate the environmental impact of a hypothetical cryptoart piece, focusing on Ethereum PoW with the current fee mechanism.
Consider an NFT with the following gas requirements:
\begin{itemize}
    \item Minting: 450k gas
    \item Buying: 300k gas
    \item Transferring: 80k gas
    \item Bidding: 100k gas
\end{itemize}

Refer to table \ref{tab:hypotheticalArt} for an estimate of the impact of these transactions.

Suppose that after minting the artwork, the creator auctions it on-chain, receiving 10 bids. The winning bidder then purchases the NFT and transfers it to a secondary account.
The impact of this art piece is therefore equal to the sum of the impacts of 1 mint, 10 bids, 1 purchase and 1 transfer. In other words, the impact of our hypothetical NFT is 467.29 kgCO\textsubscript{2}eq, equivalent to driving a typical passenger vehicle 1157 miles (or 1862 km) \citep{epa2018greenhouse}.

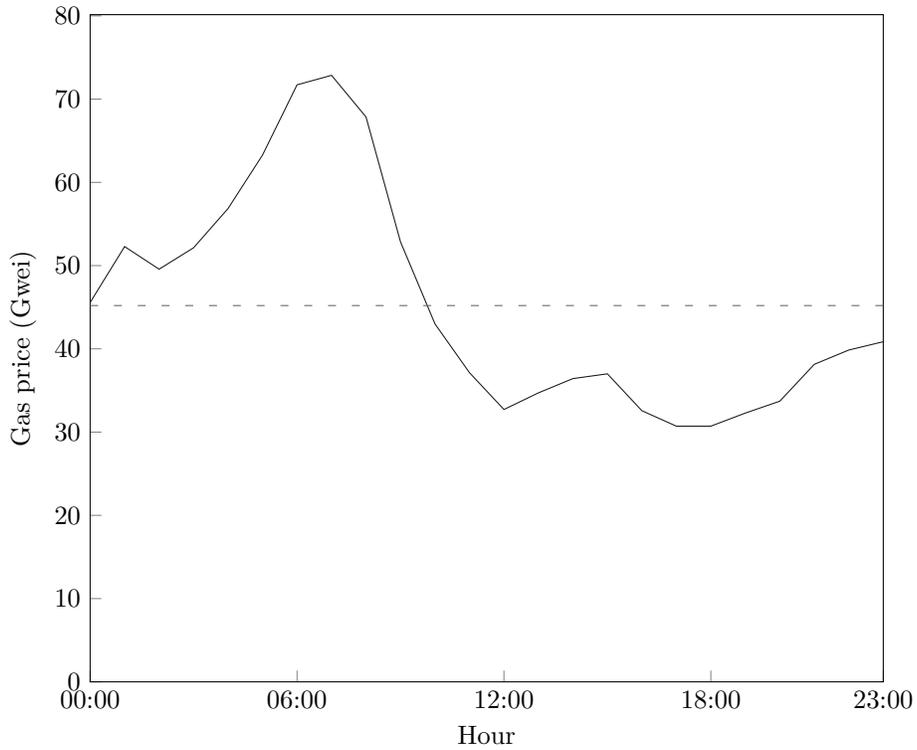
\begin{figure}[!tbhp]
    \centering
    \begin{tikzpicture}
    	\begin{axis}[width=\linewidth,
    	date coordinates in=x,
    	xticklabel={\hour:\minute},
    	xmin=2021-05-01 00:00,
    	xmax=2021-05-01 23:00,
    	ymin=0,
    	x tick label style={/pgf/number format/.cd,%
    	set thousands separator={}
    	},
    	xtick={2021-05-01 00:00, 2021-05-01 06:00, 2021-05-01 12:00, 2021-05-01 18:00, 2021-05-01 23:00},
    	xlabel = {Hour},
    	ylabel = {Gas price (Gwei}),
    	tick pos=left
    	]
    	\addplot [draw=black!90!white] table [x=Hour, y=Gas Price, col sep = semicolon] {GasPrice.csv};
    	\addplot[draw=black!60!white, loosely dashed] coordinates {(2021-05-01 00:00, 45.20) (2021-05-01 23:00, 45.20)};
    	\end{axis}
    \end{tikzpicture}
    \caption{Minimum gas price required to be included in a block by time of the day, with daily average. The data was collected from April 29, 2021 to May 5, 2021. All times are in UTC. Source: \citet{gasnow2021average}.}
    \label{fig:hourlyGasPrice}
\end{figure}

There are several simple actions that can be taken to reduce this figure:
\begin{itemize}
    \item Using off-chain bids would reduce emissions by 55\%;
    \item Executing transactions using the minimum required gas price (which, during April 29, 2021 - May 5, 2021, was on average 45 Gwei \citep{gasnow2021average}) would reduce transaction fees, and therefore emissions, by 26\%;
    \item Similarly, executing these transactions during times of the day with a lower minimum gas price, such as at 5 p.m. UTC (see table \ref{fig:hourlyGasPrice}), would reduce emissions by an additional 31\%;
\end{itemize}

All these actions together would reduce emissions to 107.37 kgCO\textsubscript{2}eq (-77\%), equivalent to driving 265 miles (or 426 km). Assuming 0.004 \$/kgCO\textsubscript{2}eq \citep{savingnature2021offset}, offsetting these emissions would cost approximately 0.43 dollars.

\begin{table}[]
\centering
\begin{tabular}{|l|l|l|l|}
\hline
\textbf{Transaction} & \textbf{Gas usage} & \textbf{Cost (USD)} & \textbf{Impact (kgCO\textsubscript{2}eq)} \\ \hline
Minting              & 450k               & 88.03               & 114.90          \\ \hline
Buying               & 300k               & 58.69               & 76.60           \\ \hline
Transferring         & 80k                & 15.65               & 20.42           \\ \hline
Bidding              & 100k               & 19.56               & 25.54           \\ \hline
\end{tabular}
\caption{Impact of different transactions of a hypothetical cryptoart piece. We use the average gas price from April 29, 2021 to May 5, 2021 of 61 Gwei \citep{etherscan2021average}, as well as the average Ether price of 3207 USD in the same period \citep{ycharts2021ethereum}.}
\label{tab:hypotheticalArt}
\end{table}

\section{Key Takeaways}
\label{sec:keyTakeaways}




In this section, we list the most meaningful results of our study.

\begin{enumerate}
    \item The impact of a transaction does not depend only on the gas used, but is instead proportional to the transaction fee;
    \item Spending 1 dollar on transaction fees on Ethereum is equivalent to emitting at least 1.305 kgCO\textsubscript{2}eq;
    \item A blockchain with full blocks pollutes more than one with empty blocks;
    \item The single most effective action that a user can take to reduce their own impact is to pay a reasonably lower gas price;
    \item Variations in the price of Ether influence the greenhouse gas emissions of the Ethereum network;
    \item As of early May 2021, Ethereum emits 28.06 MtCO\textsubscript{2}eq per year;
    \item EIP-1559 is expected to reduce global Ethereum emissions by about a third;
    \item Using Layer 2 technologies significantly reduces the impact of transactions;
    \item Transitioning to Proof of Stake remains the best long-term solution for Ethereum.
\end{enumerate}


\section{Conclusion}

Our figure of 28.06 MtCO\textsubscript{2}eq per year represents a grim remainder of the need for scalable solutions to reduce Ethereum emissions.

In this paper, we developed a model that can be used by cryptoartists to not only estimate the emissions of the entire network, but also measure their own environmental impact.
We found that, in the short term, the most effective way for a user to reduce their emissions is to execute transactions with longer waiting times or during low-activity hours.

Our model, while designed with cryptoart in mind, can be easily applied to other areas, such as decentralized finance or games. The focus on transactions allows researches and members of the art community to evaluate the effectiveness of new solutions (both on and off chain), providing an immediate feedback and highlighting which research directions are worth pursuing.

A key limitation of our model is that it only takes into account the main factors that influence the behaviour of miners. Future studies might expand upon our work by considering other costs (e.g. human labour) or sources of revenue (e.g. Miner-Extractable Value \citep{daian2020flash}).
Since the GHG emissions of the Ethereum blockchain also depend on the price of Ether, modelling the role of activities that influence its value (e.g. financial speculation) could provide a more complete picture of their environmental impact.

We hope that this paper will provide a basis for further developments in the field, with the eventual goal of building decentralized art platforms that unite artist empowerment and sustainable emissions.

\section*{Acknowledgements}
We would like to thank Davide Evangelista, Violetta Van Veen and Giuseppe Pignataro for their invaluable help with this paper.

\textit{Potential conflicts of interest}. At the time of writing, both authors own Ether and Ethereum-related cryptocurrencies. Moreover, both authors own NFTs and have been involved in the creation of NFTs and NFT platforms.

\bibliographystyle{plainnat}
\bibliography{main}

\begin{thebibliography}{44}
\providecommand{\natexlab}[1]{#1}
\providecommand{\url}[1]{\texttt{#1}}
\expandafter\ifx\csname urlstyle\endcsname\relax
  \providecommand{\doi}[1]{doi: #1}\else
  \providecommand{\doi}{doi: \begingroup \urlstyle{rm}\Url}\fi

\bibitem[Akten(2020)]{akten2020unreasonable}
Memo Akten.
\newblock {The Unreasonable Ecological Cost of \#CryptoArt}.
\newblock December 2020.
\newblock URL
  \url{https://memoakten.medium.com/the-unreasonable-ecological-cost-of-cryptoart-2221d3eb2053}.

\bibitem[AMD(2018)]{amd2018rx}
AMD.
\newblock {RX 590 specifications}, 2018.
\newblock URL \url{https://www.amd.com/en/products/graphics/radeon-rx-590}.
\newblock [Online; accessed on May 1, 2021].

\bibitem[Ardente and Talens~Peiró(2015)]{ardente2015environmental}
Fulvio Ardente and Laura Talens~Peiró.
\newblock {Environmental footprint and material efficiency support for product
  policy. Analysis of material efficiency requirements of enterprise servers}.
\newblock 2015.
\newblock \doi{http://dx.doi.org/10.2788/409022}.

\bibitem[Bentov et~al.(2016)Bentov, Gabizon, and
  Mizrahi]{bentov2016cryptocurrencies}
Iddo Bentov, Ariel Gabizon, and Alex Mizrahi.
\newblock Cryptocurrencies without proof of work.
\newblock In \emph{International conference on financial cryptography and data
  security}, pages 142--157. Springer, 2016.

\bibitem[Blockchair(2021)]{blockchair2021ethereum}
Blockchair.
\newblock {Ethereum Total transaction Chart}, 2021.
\newblock URL
  \url{https://blockchair.com/ethereum/charts/total-transaction-count}.
\newblock [Online; accessed on May 3, 2021].

\bibitem[Buterin(2021)]{buterin2021incomplete}
Vitalik Buterin.
\newblock {An Incomplete Guide to Rollups}.
\newblock January 2021.

\bibitem[CamelCamelCamel(2021)]{camel2021rx}
CamelCamelCamel.
\newblock {RX 590 price chart}, 2021.
\newblock URL \url{https://camelcamelcamel.com/product/B07PQJ2S5G}.
\newblock [Online; accessed on May 1, 2021].

\bibitem[CoinGecko(2021)]{coingecko2021ethereum}
CoinGecko.
\newblock {Ethereum (ETH)}, 2021.
\newblock URL \url{https://www.coingecko.com/en/coins/ethereum}.
\newblock [Online; accessed on May 6, 2021].

\bibitem[Daian et~al.(2020)Daian, Goldfeder, Kell, Li, Zhao, Bentov,
  Breidenbach, and Juels]{daian2020flash}
Philip Daian, Steven Goldfeder, Tyler Kell, Yunqi Li, Xueyuan Zhao, Iddo
  Bentov, Lorenz Breidenbach, and Ari Juels.
\newblock {Flash boys 2.0: Frontrunning in decentralized exchanges, miner
  extractable value, and consensus instability}.
\newblock In \emph{2020 IEEE Symposium on Security and Privacy (SP)}, pages
  910--927. IEEE, 2020.

\bibitem[De~Vries(2018)]{vries2018bitcoin}
Alex De~Vries.
\newblock {Bitcoin's growing energy problem}.
\newblock \emph{Joule}, 2\penalty0 (5):\penalty0 801--805, 2018.

\bibitem[De~Vries(2021)]{vries2021ethereum}
Alex De~Vries.
\newblock {Ethereum Energy Consumption Index}, 2021.
\newblock URL \url{https://digiconomist.net/ethereum-energy-consumption}.
\newblock [Online; accessed on May 2, 2021].

\bibitem[EIA(2021)]{eia2021monthly}
EIA.
\newblock {Monthly Electric Power Industry Report}.
\newblock \emph{U.S. Energy Information Administration}, February 2021.

\bibitem[EPA(2018)]{epa2018greenhouse}
EPA.
\newblock {Greenhouse Gas Emissions from a Typical Passenger Vehicle}.
\newblock \emph{Office of Transportation and Air Quality}, March 2018.
\newblock URL \url{https://nepis.epa.gov/Exe/ZyPDF.cgi?Dockey=P100U8YT.pdf}.

\bibitem[ESCAP(2018)]{escap2018sustainable}
ESCAP.
\newblock {Sustainable Energy in Asia and the Pacific}.
\newblock \emph{United Nations Economic and Social Commission for Asia and the
  Pacific}, 2018.
\newblock URL
  \url{https://www.unescap.org/sites/default/files/Statistical\%20Perspective\%202018\%20WEB.pdf}.

\bibitem[{EtherChain}(2021{\natexlab{a}})]{etherchain2021mining}
{EtherChain}.
\newblock {Evolution of the Total Daily Mining Reward}, 2021{\natexlab{a}}.
\newblock URL \url{https://etherchain.org/charts/miningReward}.
\newblock [Online; accessed on May 2, 2021].

\bibitem[{EtherChain}(2021{\natexlab{b}})]{etherchain2021value}
{EtherChain}.
\newblock {Evolution of the Ethereum Price in USD}, 2021{\natexlab{b}}.
\newblock URL \url{https://etherchain.org/charts/price_USD}.
\newblock [Online; accessed on May 6, 2021].

\bibitem[Etherscan(2021{\natexlab{a}})]{etherscan2021average}
Etherscan.
\newblock {Ethereum Average Gas Price Chart}, 2021{\natexlab{a}}.
\newblock URL \url{https://etherscan.io/chart/gasprice}.
\newblock [Online; accessed on May 6, 2021].

\bibitem[Etherscan(2021{\natexlab{b}})]{etherscan2021daily}
Etherscan.
\newblock {Ethereum Daily Block Rewards Chart}, 2021{\natexlab{b}}.
\newblock URL \url{https://etherscan.io/chart/blockreward}.
\newblock [Online; accessed on May 6, 2021].

\bibitem[Etherscan(2021{\natexlab{c}})]{etherscan2021ethereum}
Etherscan.
\newblock {Ethereum Network Pending Transactions Chart}, 2021{\natexlab{c}}.
\newblock URL \url{https://etherscan.io/chart/pendingtx}.
\newblock [Online; accessed on May 3, 2021].

\bibitem[Etherscan(2021{\natexlab{d}})]{etherscan2021transactions}
Etherscan.
\newblock {Transactions}, 2021{\natexlab{d}}.
\newblock URL \url{https://etherscan.io/txs}.
\newblock [Online; accessed on May 4, 2021].

\bibitem[Fairley(2019)]{fairley2019ethereum}
Peter Fairley.
\newblock {Ethereum Plans to Cut Its Absurd Energy Consumption by 99 Percent}.
\newblock \emph{IEEE Spectrum}, January 2019.
\newblock URL
  \url{https://spectrum.ieee.org/computing/networks/ethereum-plans-to-cut-its-absurd-energy-consumption-by-99-percent}.

\bibitem[GasNow(2021)]{gasnow2021average}
GasNow.
\newblock {Average GasPrice}, 2021.
\newblock URL \url{https://www.gasnow.org}.
\newblock [Online; accessed on May 3, 2021].

\bibitem[Hayes(2017)]{hayes2017cryptocurrency}
Adam~S Hayes.
\newblock {Cryptocurrency value formation: An empirical study leading to a cost
  of production model for valuing bitcoin}.
\newblock \emph{Telematics and Informatics}, 34\penalty0 (7):\penalty0
  1308--1321, 2017.

\bibitem[Howson et~al.(2019)Howson, Oakes, Baynham-Herd, and
  Swords]{howson2019cryptocarbon}
Peter Howson, Sarah Oakes, Zachary Baynham-Herd, and Jon Swords.
\newblock Cryptocarbon: the promises and pitfalls of forest protection on a
  blockchain.
\newblock \emph{Geoforum}, 100:\penalty0 1--9, 2019.

\bibitem[IEA(2020)]{iea2020composition}
IEA.
\newblock {Composition of CO2 emissions and emission intensity in 2020}, 2020.
\newblock URL
  \url{https://www.iea.org/data-and-statistics/charts/composition-of-co2-emissions-and-emission-intensity-in-2020}.

\bibitem[Immutable(2020)]{immutable2020immutable}
Immutable.
\newblock {Immutable X is Making NFTs Carbon Neutral on Ethereum}.
\newblock October 2020.
\newblock URL \url{https://www.immutable.com/blog/carbon-neutral-nfts}.

\bibitem[Kahn(2021)]{kahn2021how}
Brian Kahn.
\newblock {How to Fix Crypto Art NFTs' Carbon Pollution Problem}.
\newblock March 2021.
\newblock URL
  \url{https://earther.gizmodo.com/how-to-fix-crypto-art-nfts-carbon-pollution-problem-1846440312}.

\bibitem[Kanani et~al.(2019)Kanani, Nailwal, and Arjun]{kanani2019matic}
Jaynti Kanani, Sandeep Nailwal, and Anurag Arjun.
\newblock {Matic Whitepaper}.
\newblock 2019.

\bibitem[KnownOrigin(2021)]{knownorigin2021step}
KnownOrigin.
\newblock {A step toward sustainability}.
\newblock 2021.
\newblock URL
  \url{https://knownorigin.io/journal/platformupdate/a-step-toward-sustainability}.

\bibitem[Lemercier(2021)]{lemercier2021joanie}
Joanie Lemercier.
\newblock {The problem of CryptoArt}.
\newblock February 2021.
\newblock URL \url{https://joanielemercier.com/the-problem-of-cryptoart/}.

\bibitem[Lin et~al.(2021)Lin, Li, Zhao, and Chen]{lin2021measuring}
Qinwei Lin, Chao Li, Xifeng Zhao, and Xianhai Chen.
\newblock {Measuring decentralization in bitcoin and ethereum using multiple
  metrics and granularities}.
\newblock \emph{arXiv preprint arXiv:2101.10699}, 2021.

\bibitem[Mattei(2021)]{mattei2021should}
Shanti Escalante-De Mattei.
\newblock {Should You Worry About the Environmental Impact of Your NFTs?}
\newblock \emph{ARTNews}, April 2021.
\newblock URL
  \url{https://www.artnews.com/art-news/news/nft-carbon-environmental-impact-1234589742/}.

\bibitem[Minerstat(2021)]{minerstat2021gpu}
Minerstat.
\newblock {GPU Mining Leaderboard}, 2021.
\newblock URL \url{https://minerstat.com/hardware/gpus?algo=Ethash}.
\newblock [Online; accessed on May 1, 2021].

\bibitem[Pipkin(2021)]{pipkin2021here}
Everest Pipkin.
\newblock {Here is the article you can send to people when they say “But the
  environmental issures with cryptoart will be solved soon, right?”}.
\newblock March 2021.
\newblock URL
  \url{https://everestpipkin.medium.com/but-the-environmental-issues-with-cryptoart-1128ef72e6a3}.

\bibitem[Poon and Buterin(2017)]{poon2017plasma}
Joseph Poon and Vitalik Buterin.
\newblock {Plasma: Scalable autonomous smart contracts}.
\newblock \emph{White paper}, pages 1--47, 2017.

\bibitem[Rademaekers et~al.(2020)Rademaekers, Smith, and
  Torres~Vega]{rademaekers2020study}
Koen Rademaekers, Matthew Smith, and Perla C. et~al. Torres~Vega.
\newblock {Study on energy prices, costs and their impact on industry and
  households - Final Report}.
\newblock 2020.
\newblock URL
  \url{https://op.europa.eu/en/publication-detail/-/publication/16e7f212-0dc5-11eb-bc07-01aa75ed71a1/language-en}.

\bibitem[Roughgarden(2020)]{roughgarden2020transaction}
Tim Roughgarden.
\newblock {Transaction Fee Mechanism Design for the Ethereum Blockchain: An
  Economic Analysis of EIP-1559}.
\newblock \emph{arXiv preprint arXiv:2012.00854}, 2020.

\bibitem[SavingNature(2021)]{savingnature2021offset}
SavingNature.
\newblock {Carbon Footprint Calculator}, 2021.
\newblock URL
  \url{https://savingnature.com/offset-your-carbon-footprint-carbon-calculator/}.
\newblock [Online; accessed on May 7, 2021].

\bibitem[Silva et~al.(2020)Silva, Vavricka, Barreto, and
  Matos]{silva2020impact}
Paulo Silva, David Vavricka, Joao Barreto, and Miguel Matos.
\newblock {Impact of geo-distribution and mining pools on blockchains: a study
  of Ethereum}.
\newblock In \emph{2020 50th Annual IEEE/IFIP International Conference on
  Dependable Systems and Networks (DSN)}, pages 245--252. IEEE, 2020.

\bibitem[SuperRare(2021)]{superrare2021no}
SuperRare.
\newblock {No, CryptoArtists Aren’t Harming the Planet}.
\newblock 2021.
\newblock URL
  \url{https://medium.com/superrare/no-cryptoartists-arent-harming-the-planet-43182f72fc61}.
\newblock [Online; accessed on May 3, 2021].

\bibitem[Taylor(2017)]{taylor2017evolution}
Michael~Bedford Taylor.
\newblock The evolution of bitcoin hardware.
\newblock \emph{Computer}, 50\penalty0 (9):\penalty0 58--66, 2017.

\bibitem[Wintermeyer(2021)]{wintermeyer2021climate}
Lawrence Wintermeyer.
\newblock {Climate-Positive Crypto Art: The Next Big Thing Or NFT Overreach?}
\newblock \emph{Forbes}, March 2021.
\newblock URL
  \url{https://www.forbes.com/sites/lawrencewintermeyer/2021/03/19/climate-positive-crypto-art-the-next-big-thing-or-nft-overreach/?sh=482cfa04b0e6}.

\bibitem[Wood et~al.(2014)]{wood2014ethereum}
Gavin Wood et~al.
\newblock {Ethereum: A secure decentralised generalised transaction ledger}.
\newblock \emph{Ethereum project yellow paper}, 151\penalty0 (2014):\penalty0
  1--32, 2014.

\bibitem[YCharts(2021)]{ycharts2021ethereum}
YCharts.
\newblock {Ethereum Price}, 2021.
\newblock URL \url{https://ycharts.com/indicators/ethereum_price}.

\end{thebibliography}

\appendix
\section{Closed-Form Expression of $R_t$}
\label{app:closedForm}

Let $\psi_q(x) = \frac{d}{dx} \Gamma_q (x)$ be the $q$-digamma function.
Let
\begin{equation}
    h = \frac{\log\left(\frac{m}{m+KS_0-S_0}\right)}{\log K}
\end{equation}
The closed-form expression of $R_t$ is:
\begingroup
\begin{equation}
    mV\left(\frac{(K-1)\psi_K\left(t-h+1\right)}{m\log K}+\frac{(1-K)\psi_K\left(-h\right)}{m\log K}-\frac{(K-1)(t+1)}{m}\right)
\end{equation}
\endgroup

where $\log$ is the natural logarithm.

\begin{equation}
    \Delta S_t = m - \frac{b}{V} S_t
\end{equation}

\begin{equation}
    \frac{dS}{dt} = m - \frac{b}{V} S
\end{equation}

\begin{equation}
    S_t = c_1 e^{-bt/V} + \frac{mV}{b}
\end{equation}
For $t = 0$, the formula is equal to the initial circulating supply $S_0$. Therefore:
\begin{equation}
    c_1 = S_0 - \frac{mV}{b}
\end{equation}
\begin{equation}
    S_t = \frac{mV}{b} \left (S_0 - e^{-bt/V} \right )
\end{equation}

\end{document}